\documentclass[conference]{IEEEtran}
\IEEEoverridecommandlockouts

\usepackage[utf8]{inputenc}

\usepackage[sorting=none,style=ieee]{biblatex}     %
\def\BibTeX{{\rm B\kern-.05em{\sc i\kern-.025em b}\kern-.08em T\kern-.1667em\lower.7ex\hbox{E}\kern-.125emX}}

\usepackage[acronym, nowarn]{glossaries}
\makeglossaries
\newacronym{mmtc}{mMTC}{massive machine type communication}
\newacronym{mac}{MAC}{multiple access control}
\newacronym{bs}{BS}{base station}
\newacronym{ue}{UEs}{user equipments}
\newacronym{ctu}{CTU}{contention transmission unit}
\newacronym{scma}{SCMA}{sparse code multiple access}
\newacronym{ofdma}{OFDMA}{orthogonal frequency-division multiple access}
\newacronym{noma}{NOMA}{non-orthogonal multiple access}
\newacronym{rl}{RL}{reinforcement learning}
\newacronym{eab}{EAB}{enhance access bearing}
\newacronym{qmix}{QMIX}{Monotonic Value Function Factorisation for Deep Multi-Agent Reinforcement Learning}
\newacronym{rr}{RR}{Round Robin}
\newacronym{wf}{WF}{Water Filling}
\newacronym{wflb}{WFLB}{Water Filling Lower Bound}
\newacronym{drl}{DRL}{deep reinforcement learning}
\newacronym{madrl}{MADRL}{multi-agent deep reinforcement learning}
\newacronym{lstm}{LSTM}{long short-term memory}
\newacronym{dqn}{DQN}{Deep Q Networks}
\newacronym{tdd}{TDD}{time duplex division}
\newacronym{cw}{CW}{contention window}
\newacronym{csi}{CSI}{channel state information}
\newacronym{urllc}{URLLC}{ultra-realiable and low-latency communication}
\newacronym{decpomdp}{Dec-POMDP}{decentralized partial observable Markov decision process}
\newacronym{dnn}{DNN}{deep neural networks}
\newacronym{ddql}{DQL}{distributed Q-learning}
\newacronym{iot}{IoT}{Internet of Things}
\newacronym{gf}{GF}{grant-free}
\newacronym{qlbt}{QLBT}{QMIX Listen Before Talk}
\newacronym{flops}{FLOPS}{FLoating-point Oerations Per Second}

\usepackage{amsmath, amsthm, amssymb, amsfonts, physics, upgreek}

\usepackage{graphicx}
\usepackage{subcaption}

\usepackage{algorithm} 
\usepackage{algpseudocode}
\algnewcommand{\Initialize}[1]{%
  \State \textbf{Initialize:}
  \Statex \hspace*{\algorithmicindent}\parbox[t]{.8\linewidth}{\raggedright #1}
}

\usepackage{booktabs}
\usepackage{multirow}
\usepackage[table]{xcolor}

\usepackage{soul}
\usepackage{color}

\usepackage{hyperref}

\begin{document}

\title{TinyQMIX: Distributed Access Control for mMTC via Multi-agent Reinforcement Learning
}

\author{
    \IEEEauthorblockN{
        Tien Thanh Le\IEEEauthorrefmark{1}\IEEEauthorrefmark{2}, 
        Yusheng Ji\IEEEauthorrefmark{2}\IEEEauthorrefmark{1}, 
        John C.S. Lui \IEEEauthorrefmark{3}
    }
    \IEEEauthorblockA{
        \IEEEauthorrefmark{1}Department of Informatics, The Graduate University for Advanced Studies, SOKENDAI, Tokyo, Japan\\
        \IEEEauthorrefmark{2}National Institute of Informatics, Tokyo, Japan\\
        \IEEEauthorrefmark{3}Department of Computer Science and Engineering, The Chinese University of Hong Kong, Hong Kong\\
        Email: \IEEEauthorrefmark{1}\IEEEauthorrefmark{2}\{lethanh, kei\}@nii.ac.jp;\IEEEauthorrefmark{3}cslui@cse.cuhk.edu.hk
    }
}

\maketitle

\begin{abstract}
Distributed access control is a crucial component for \gls{mmtc}.
In this communication scenario, centralized resource allocation is not scalable because resource configurations have to be sent frequently from the base station to a massive number of devices.
We investigate distributed reinforcement learning for resource selection without relying on centralized control.
Another important feature of \gls{mmtc} is the sporadic and dynamic change of traffic.
Existing studies on distributed access control assume that traffic load is static or they are able to gradually adapt to the dynamic traffic.
We minimize the adaptation period by training TinyQMIX, which is a lightweight multi-agent deep reinforcement learning model, to learn a distributed wireless resource selection policy under various traffic patterns before deployment.
Therefore, the trained agents are able to quickly adapt to dynamic traffic and provide low access delay.
Numerical results are presented to support our claims.
\end{abstract}
\glsresetall

    \section{Introduction}

In recent years, the number of connected devices has grown exponentially due to the proliferation of \gls{iot} applications.
Many \gls{iot} applications are enabled by \gls{mmtc}, i.e, autonomous vehicles, industrial automation, or environmental sensing.
It has been projected that about half of total global connections (about 15 billion devices) will be \gls{mmtc} devices \cite{shahab2020grant}.

The traffic features of mMTC are inherently different from other communication scenarios.
First, \gls{mmtc} protocols should support a \emph{high overloading factor}, in which a large number of devices share a small number of wireless resources.
Although the number of wireless resources is small, the limited resource would still be able to support \gls{mmtc} because \gls{mmtc} devices typically transmit a \emph{low volume of short packets and uplink data} \cite{navarro2020survey}.
Another important feature of \gls{mmtc} is \emph{sporadic traffic} \cite{chen2020massive}.
Sporadic traffic means that devices do not always have data to transmit and only a random subset of devices access the network in each timeslot.
Moreover, \gls{mmtc} traffic tends to be \emph{dynamic}, which means that the amount of data generated and sent by \gls{mmtc} devices would unlikely be a constant rate throughout.
For example, \gls{iot} sensors normally send regular status updates to the server, but sometime an important event would be triggered, so these devices must send a larger amount of information regarding that important event \cite{navarro2020survey}.
In short, future \gls{mmtc} protocols are expected to address the mentioned features, by allowing high overloading, and supporting sporadic, and dynamic uplink traffic.


To solve this problem, many of approaches have been put forward in 5G's \gls{mac} layer such as uplink contention-based \gls{gf}-\gls{noma} \cite{au2014uplink}.
\gls{noma} would increase the system overloading because it enables multiple devices to reuse the same time-frequency resource via power domain or code domain multiplexing.
The contention-based grant-free mechanism reduces the delay of sporadically arrived packets since it allows devices to transmit data directly to \gls{bs} without the need for the request and grant procedure.

A growing body of work has improved the original contention-based \gls{gf} \gls{noma} by equipping the protocol with reinforcement learning.
In general, reinforcement learning techniques can solve the network optimization problem in a data-driven manner instead of relying on the analytical network model.
For the current problem, reinforcement learning is adopted for selecting the \gls{mac} layer's parameters.
In \cite{zhang2020deep}, \gls{dqn} with \gls{lstm} architecture are deployed at each \gls{mmtc} device to select the wireless resource and power level to maximize the network throughput.
This is an independent \gls{dqn} policy.
However, for independent learning methods, the changing policy of one agent leads to changing the optimal target policy of another agent.
Thus, \emph{independent \gls{dqn} has no convergence guarantees and may lead to poor performance} \cite[Sec 3.3]{rashid2018qmix}.

The problem of independent \gls{dqn} can be mitigated by \gls{madrl} policies such as \gls{qmix} \cite{rashid2018qmix}.
In \gls{qmix}, the \gls{lstm} Q-function of all devices are trained together in a centralized simulator, such that the Q-value of an agent is consistent with the joint-action's Q-value function of all agents.
After training, agents can select the best joint-action independently without any communication.
\citeauthor{huang2020throughput} leveraged \gls{qmix} for selecting pilot sequences \cite{huang2020throughput}, while \citeauthor{guo2022multi} also applied \gls{qmix} for deciding whether to back-off in 802.11 network \cite{guo2022multi}.
Both proposed \gls{qmix} policies demonstrated their superiority to independent \gls{dqn} and other heuristic policies.
However, the main drawback of these policies is \emph{the computational complexity of their \gls{lstm} architecture}. In fact, running this neural network's architecture on \gls{iot} devices take a long time while the deadline for decision-making is short.
Also, the problem of \emph{low rate and sporadic traffic were not explicitly addressed}.

Sporadic traffic for uplink contention-based \gls{noma} has been investigated in \cite{liu2021distributed}.
two \gls{ddql} schemes denoted as ADDQ and PDDQ to select resources that minimize collisions.
However, they only consider that \emph{traffic intensity of each device remains constant throughout its lifetime}.

To the best of our knowledge, there is no prior work on a fully distributed \gls{mac} protocol that addresses the wireless resource selection problem with sporadic and dynamic traffic in \gls{mmtc}.
Therefore, the goal of this work is to design a new distributed resource selection protocol to solve this problem.
The primary contributions of this paper are as follows:
\begin{itemize}
    \item We map the problem into a \gls{decpomdp}. We also design a lightweight local observation and \gls{madrl} policy called TinyQMIX. TinyQMIX agents have a low model complexity, such that they can be implemented on \gls{mmtc} devices
    \item TinyQMIX policy is trained on a wide variety of sporadic traffic scenarios, allowing the trained agents to quickly adapt to the ever-changing traffic dynamic. Devices can independently select the best wireless resource to minimize system-wide access delay.
    \item Numerical simulations are performed to evaluate the transmission delay of the TinyQMIX with other widely-adopted heuristic and \gls{madrl} algorithms.
\end{itemize}

The rest of this paper is organized as follows.
Section \ref{sec:system} presents the network and traffic model.
We propose our \gls{decpomdp} formulation and TinyQMIX for dynamic access control in \gls{mmtc} in the Section \ref{sec:propose}.
Numerical simulation is presented in Section \ref{sec:simulation}
    \section{System Model}
\label{sec:system}
Here, we present the scope of our study, which includes the \gls{mmtc} network model, the dynamic and sporadic traffic model, and the dynamic access control model.

\subsection{Network scenario}

\begin{figure}
    \centering
    \includegraphics[width=\linewidth]{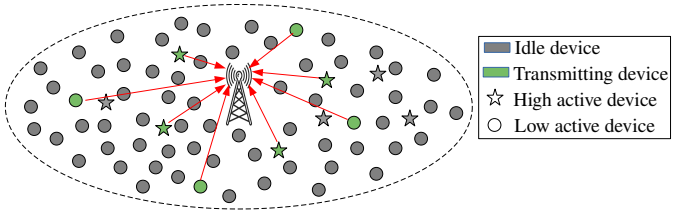}
    \caption{\gls{mmtc} devices uplink transmission model}
    \label{fig:model}
\end{figure}
For the ease of presentation, we first consider a single-cell wireless system in Figure \ref{fig:model}. 
The \gls{bs} is located at the center and is surrounded by $N$ devices within the coverage radius.
Let $\mathcal{N}=\{1, 2, \ldots, i, \ldots, N\}$ be the set of $N$ devices.
Assume that the system is slotted and time synchronized.
Because the \gls{mmtc} traffic contains mostly short packets, we assume that the service time of one packet is one timeslot.

\subsection{Traffic model}

Assume that the packet arrival rate per timeslot for the $i^{th}$ device is $\lambda_i(t)$ where $t$ is the index of a timeslot.
Similar to \cite{liu2021distributed}, we assume the set of devices $\mathcal{N}$ can be divided into two groups: high active devices $\mathcal{N}_h$ and low active devices $\mathcal{N}_l$.
Also, the network is likely to contain more low active devices ($|\mathcal{N}_h| \ll |\mathcal{N}_l|$).

We consider the dynamic traffic arrival scenario: The distribution of traffic of every device changes within their lifetime.
Therefore, the parameter of packet arrival distribution $\lambda(t)$ is a time-dependent variable.
We assume that devices have the same dynamic within an interval $\Delta T$, where $\Delta T$ is a constant.
All devices change their arrival distribution $\lambda(t)$ synchronously after a constant $\Delta T$ timeslots.
We like to note that since our proposed method is data-driven, it would also work with asynchronous changes.
\subsection{Distributed access control model}
\label{sec:system:dacm}

Let the total duration of $\tau$ timeslots be the scheduling interval.
At the beginning of every scheduling interval, each device selects any of $C$ resource units to send data.
Thus, the scheduling decision space is $C^N$, which grows exponentially with the number of devices and resources.
We reduce the scheduling space by dividing $N$ devices into many groups, each containing $N^{'}$ devices and $M$ resource units, similar to what has been done in \cite{liu2021distributed,jiang2018distributed,zhang2020deep}. 
Devices can be grouped arbitrarily as long as less than $N'$ devices share $M$ resources.
Let $\phi=\frac{N^{'}}{M}$ be the overloading factor, and $\phi \gg 1$.

A collision occurs when two or more devices select the same resource unit and transmit their data in the same timeslot.
Whenever this happens, each colliding device follows a binary exponential back-off procedure to resolve the contention.
Particularly, each device retains a contention window value $cw$, which is initialized at $1$.
If there is a collision, the contention window is doubled until it reaches $CW_{\text{max}}$.
Then, each device draws a uniform random integer $r \sim U([1, cw])$, and waits for $r$ timeslots before resuming the transmission.
To counter the effect of sporadic traffic, each device can temporarily store a maximum of $L_{\text{buffer}}$ packets in its buffer.
Also, a collided packet can be retried for a maximum of $L_{\text{retry}}$ times.
By doing that, the fraction of dropped packets is negligible, and the high-reliability requirement could be realized.

The distributed access control mechanism here is a general abstraction and it can be integrated with contention-based \gls{gf}-\gls{noma}.
A resource unit in \gls{gf}-\gls{noma} can be a tuple of frequency, pilot sequence, and \gls{noma} code-book.
We suppose that the given system adopts \gls{tdd} mode.
Pilot sequences are broadcasted by \gls{bs}, then devices measure \gls{csi} to calibrate their power level to satisfy the receiving power requirement.
Assume that devices are able to satisfy the requirement, so the only cause for transmission failure is transmission collision.

    \section{Distributed Dynamic Resource Selection - TinyQMIX}
\label{sec:propose}

The distributed wireless resource selection problem can be formulated as a \gls{decpomdp} \cite{oliehoek2016concise}, which is defined as a tuple $G = \langle N', S, U, P, r, Z, \gamma \rangle$.
The system consists of $N'$ agents (or $N'$ devices).
Each agent runs according to Algorithm \ref{alg:agent} with TinyQMIX decision-making capability combined with random access procedure.
$s \in S$ is the true state of the environment.
The state can either be: (1) the state of the network simulation's program when the agents are trained, (2) the state of the \gls{bs} and channel quality at deployment time.

\begin{algorithm}
    \small
        \caption{TinyQMIX agent $i$ for distributed access control}
        \begin{algorithmic}[1]
            \Require{Agent's DNN parameter $\theta^i$, MAC's parameter $CW_{\text{max}}, L_{\text{buffer}}, L_{\text{retry}}$}
            \Initialize{$\bar{\lambda}^i=0$, $u_{t-1}^i=0$, $\langle \bar{sr}^1, \ldots, \bar{sr}^M \rangle = \vb{0}$}
            \For{$t=1, 2, \ldots$}
                \State Generate packets
                \State Append packets to buffer until $L_{\text{buffer}}$ packets are stored
                \If{$t \text{ mod } \tau = 0$}
                    \State Select resource unit $u_t^i=\text{argmax}_{u^i} \; Q_i(z^i, u^i; \theta^i)$
                \EndIf
                \State Back-off or transmit the head packet in the buffer on resource unit $u_t^i$ (see Section \ref{sec:system:dacm})
                \State Receive acknowledgement packet from \gls{bs}
                \State Adjust back-off parameter $cw$ or drop packet (see Section \ref{sec:system:dacm})
                \State Update local observation $z^i$ includes $\bar{\lambda}^i$, $\langle \bar{sr}^1, \ldots, \bar{sr}^M \rangle$ using Formula \ref{eq:incremental-update}
                \State Update the running mean and variance of $z^i$ at the training phase, and normalize $z^i$ in all phases
            \EndFor
        \end{algorithmic}
        \label{alg:agent}
\end{algorithm}

In every timeslot, each device $i \in \{1, \ldots, N'\}$ chooses an action $u^i \in U$, which is a resource unit.
Here, $U$ is a group of resources, which is granted a group of devices.
The actions of all devices in that group constitute the joint action $\vb{u} = (u^1, \ldots, u^{N'}) \in \vb{U}$.
If the joint action is $\vb{u}$ under the state $s$, the next state of the system $s'$ can be obtained according to the state transition function $P(s'|s, \vb{u}): S \times \vb{U} \rightarrow \vb{S}$.
For this \gls{decpomdp}, we assume it is of a model-free structure.
During the training phase, the network simulator can generate the transition function $P$.
Given the current network state $s$ and the joint-action $\vb{u}$, the simulator computes the next state $s'$ according to the traffic model, binary exponential back-off rule, and buffer rule, which is described in Section \ref{sec:system}.
We discuss the remaining elements of the \gls{decpomdp} formulation in the following subsections.

\subsection{Measurement of local observation}

In \gls{decpomdp}, partially observable means that each of $n$ agents cannot obtain the true state $s$ because $s$ contains a network-wide information.
However, devices can extract their local observation $z \in Z$, which can be easily gathered on their own.
The local observation is the input for devices to determine the next action without communicating to the \gls{bs}.
We design the local observation as follows:

\begin{enumerate}
    \item the average packet arrival rate $\bar{\lambda}^i$
    \item the previous action $u^i_{t-1}$
    \item the list of average success transmission rate per resource $\langle \bar{sr}^1, \ldots, \bar{sr}^M \rangle$.
\end{enumerate}
The first two elements represent the internal state of each device, whereas the remaining elements capture partial information about the network's traffic intensity at each resource from the perspective of that device.

Each device selects an action based on a stochastic policy $\pi^i(u)$ (Line 5-7, Algorithm \ref{alg:agent}).
In particular, the policy is maximizing the Q-value $\text{argmax}_{u^i} \; q_i = Q_i(z^i, u^i; \theta^i)$, where $\theta^i$ is the \gls{dnn}'s parameter.
Previous works adopted a sequence of historical local observation as the input for each agent and \gls{lstm} as the network architecture \cite{zhang2020deep,huang2020throughput,guo2022multi}.
This approach demanded a large memory footprint and lengthy computation, but \gls{mmtc} devices have limited computational capability.
Thus, a compact local observation and small neural networks architecture is needed.
The local observation for agents in our proposed system is a vector of $M + 2$ elements.
Also, a fully connected \gls{dnn} with one hidden layer is the neural networks architecture.
Besides, to minimize the memory footprint on the \gls{mmtc} devices, the historical average is captured via an incremental implementation \cite[p.31]{sutton2018reinforcement}. 
For example, the update rule to estimate the average packet arrival rate is:
\begin{equation}
    \bar{\lambda}^i_{t+1} \leftarrow \bar{\lambda}^i_{t} + \alpha (x_{t} - \bar{\lambda}^i_{t})
    \label{eq:incremental-update}
\end{equation}
where $x_{t}$ is the number of packet generated within the $t^{th}$ timeslot, and $\alpha$ is the step size of the update.
A similar update rule is applied for estimating the success rates.

The learning process could be hindered if the scale of different inputs to the \gls{dnn} are not the same.
Also, agents do not know the exact scale of the average packet arrival rate as well as the average success transmission rate.
Thus, we track the running mean and variance of the local observations using the data generated during the training phase of the system \cite{knuth1997art}.
This technique allows the mean and variance to be estimated as the observation arrives one at a time, while devices do not need to keep the observation for a second pass.
Devices learn the means and variances during the training phase.
Then, the final means and variances of the observations are kept constant to normalize the observation in the testing and deployment phase (Line 12, Algorithm \ref{alg:agent}).


\subsection{Global observation and reward}

\emph{The global observation} is the concatenation of local observations $\vb{z} = \langle z_1, \ldots, z_{N'} \rangle$.
This global observation is the input for the Hypernetwork of \gls{qmix}, which generates the parameter for the mixer network.
The mixer network combines the values given by individual value functions $\vb{q} = \langle q_1, \ldots, q_{N'} \rangle$ into a single value $\hat{q}_{\text{tot}}$, which estimate the joint-action's value.
In the training phase, the gradient is backpropagated from the mixer network to individual \gls{dnn} value functions.
This mechanism allows each device to learn the association between its local observation and the global observation, thereby leading to higher performance than independent \gls{dqn}.

Let $r(s, \vb{u}): S \times \vb{U} \rightarrow \mathbb{R}$ be the joint-action reward function.
Consider cluster of devices $\mathcal{N}^{'}$ at timeslot $t$, the set of devices that attempt to transmit is $\mathcal{N}^{'}_{\text{transmit}}(t)$.
The set of successfully transmitted devices is $\mathcal{N}^{'}_{\text{success}}(t)$.
Then, the joint-action reward at every scheduling interval $\uptau$ is defined as:
\begin{equation}
    r(t) = \frac{1}{N^{'}}\sum_{i = 1}^{i=N^{'}}  \frac{\sum_{t'=t}^{t'=t+\uptau} 1\{i \in \mathcal{N}^{'}_{\text{success}}(t)\}}{\sum_{t'=t}^{t'=t+\uptau} 1\{i \in \mathcal{N}^{'}_{\text{transmit}}(t)\}}
    \label{eq:reward}
\end{equation}
Equation (\ref{eq:reward}) presents the average success transmission rate overall $N^{'}$ devices in a scheduling interval $\uptau$.
If the reward is higher, the average access delay is lower because a high success rate means that devices do not have to perform am excessive amount of back-off operations.
This reward function is similar to previous work on sporadic traffic \cite{liu2021distributed}.
The reward is chosen to be the average success rate over a longer time horizon such that the uncertainty of the sporadic traffic arrival is mitigated.

\subsection{Training with dynamic traffic}

TinyQMIX agents can be trained to select the best wireless resource under various traffic conditions according to Algorithm \ref{alg:offline-training}.
First, the \gls{dnn} for every device and the mixer network are initialized at random.
The average arrival rate is changed every interval $\Delta T$ (Line 3-5, Algorithm \ref{alg:offline-training}).
We train the TinyQMIX agents such that they can cooperatively select the resource units in an uncoordinated manner.
Particularly, the TinyQMIX parameter is optimized to estimate the correct joint-action Q-value with different sporadic traffic distributions (Line 13-16, Algorithm \ref{alg:offline-training}).

In the testing phase, we generate and store traffic traces.
A trace is a matrix containing the number of packets generated by each device in every timeslot.
Different methods are tested on the same trace to compare their performance.
When being tested on the trace, the trained TinyQMIX models do not require any further fine-tuning or adaptation.
The trained model can immediately perform well on newly unseen traffic distribution in the testing traffic traces because the \gls{dnn} agents can generalize and give an accurate assessment of the action's value based on the vast number of trained patterns.

\begin{algorithm}
    {\small
        \caption{An episode of offline centralized training}
        \begin{algorithmic}[1]
            \State $t \leftarrow 0$
            \While{$t < T_{episode}$}
                \If{$t \text{ mod } \Delta T = 0$}
                    \State Redraw $\lambda(t)$
                \EndIf
                \For{step in $1, \ldots, T_{\text{optimization interval}}$}
                    \State Run all agents (Algorithm \ref{alg:agent})
                    \State Collect local data $\vb{z}_t$, $\vb{z}_{t+1}$, $\vb{u}_t$
                    \State Collect global reward $r(t)$
                    \State Save $\langle \vb{z}_t, \vb{z}_{t+1}, \vb{u}_t, r(t) \rangle$ to replay memory
                \EndFor
                \State Sample minibatch from replay memory
                \State Compute individual's Q-values
                    $$\vb{q} = \langle Q_i(z^i, u^i; \theta^i), \forall i \in \{1, \ldots, N'\} \rangle$$ 
                \State Compute estimated total Q-value
                    $$\hat{q}_{tot} = Q_{tot}(\vb{z}, \vb{u};\theta^{\text{mixer}})(\vb{q})$$
                \State Compute target total Q-value
                    $${q}_{tot} = r + \gamma \max_{\vb{u}'}Q_{tot}(\vb{z}, \vb{u}';\theta^{\text{mixer}})$$
                \State Compute the mean square error between $\hat{q}_{tot}$ and ${q}_{tot}$. Compute the gradient using the error and update the network parameters using stochastic gradient ascend
                \State $t \leftarrow t + 1$
            \EndWhile
        \end{algorithmic}
        \label{alg:offline-training}
    }
\end{algorithm}

    \section{Numerical Simulation}
\label{sec:simulation}
In this section, we discuss the simulation scenario, briefly introduce the practical baselines, and finally provide an empirical comparison.

\subsection{Simulation scenario}
    We tested different channel access policies under three levels of traffic dynamic: $\Delta T \in \{10\text{s}, 60\text{s},\infty\}$. 
    The traffic of the set of high active devices $\mathcal{N}_h$ has Poisson distribution with average arrival rate $\lambda_h=0.1$ (packet per slot), whereas that of the low active devices $\mathcal{N}_l$ also has Poisson distribution with $\lambda_l=0.00833$ \cite{liu2021distributed}.
    The device type is randomly reassigned such that the probability of high active device is $1/5$, and the probability of low active device is $4/5$. 
    
    The total length of the traffic traces for testing is $1$ hour.
    We consider the $0.5$ ms timeslot, then the testing time is equal to $7.2$ million timeslots.
    Let the scheduling interval be $25$ms (50 timeslots), then there are a total of $144k$ scheduling intervals.
    Also, the MAC's parameters $L_{\text{buffer}}$, $L_{\text{retry}}$, and $CW_{\text{max}}$ are all equal to $16$.
    These parameters were chosen such that the probability of packet drop is negligible. 
    Thus, different policies are only compared in terms of their access delay.

    Then, we tested the system with different cluster sizes.
    The number of resource units per cluster are $M \in \{2, 4, 8, 16\}$, and the number of devices per cluster are $N' \in \{12, 24, 48, 96\}$, respectively.
    The system overloading is $\phi = 6$.
    Similar to \cite{liu2021distributed}, the ratio between high and low active devices is $\frac{|\mathcal{N}_h|}{|\mathcal{N}_l|} = \frac{1}{4}$.
    The detailed hyperparameters for training TinyQMIX are presented in Table \ref{tab:training:param}.

    \begin{table}[h]
    \centering
    \caption{Hyperparameters for training TinyQMIX}
    \begin{tabular}{cc}
    \hline
    \rowcolor{gray!50}
    \textbf{Hyper parameters} & \textbf{Value} \\ \hline
    Number of training episodes & 1000 \\
    \rowcolor{gray!25}
    Episode length & 100(s) \\
    Optimization interval & 32 \\
    \rowcolor{gray!25}
    Learning rate & 1e-4 \\ 
    Batch size & 1024 \\ 
    \rowcolor{gray!25}
    Replay memory size & 10000 \\ 
    Discounted factor $\gamma$ & 0 \\ 
    \rowcolor{gray!25}
    Exploration start $\epsilon_{\text{start}}$ & 0.9 \\
    Exploration end $\epsilon_{\text{end}}$ & 0.05 \\
    \rowcolor{gray!25}
    Number of agents & \{12, 24, 48, 96\} \\
    Observation update's step size $\alpha$ & 0.001 \\
    \rowcolor{gray!25}
    Number of hidden units for agents' \gls{dnn} & \{8, 8, 16, 32\} \\ 
    Number of hidden units for mixer's \gls{dnn} & \{64, 128, 256, 512\} \\
    \hline
    \label{tab:training:param}
    \end{tabular}
    \end{table}

\subsection{Baselines}
We compare TinyQMIX policy with widely-used distributed resource allocation policies.
First, random is a policy that each device selects the action randomly.
Second, \gls{rr} is a policy in which each device takes turn for transmitting sequentially in each available resource unit and sequentially over time.
\gls{ddql} which has been proposed in \cite{liu2021distributed} is also taken into account.
We also compare our proposal with recently proposed QMIX policy using LSTM as agent's network architecture \cite{huang2020throughput,guo2022multi}, denotes LSTMQMIX.
Similar to what has been done in \cite{guo2022multi}, we kept a sequence of 10 local observations as the input for the LSTM agents. 
Then, we trained the LSTMQMIX policy using Algorithm \ref{alg:offline-training}.

\gls{wf} is a heuristic and centralized policy, in which, the \gls{bs} can estimate the average arrival rate to allocate a balanced amount of traffic on different resource units.
However, in \gls{wf}, devices must send the \gls{csi} to \gls{bs}, then \gls{bs} must take a proportion of timeslots to send the resource assignment to devices.
\gls{wflb} is an unrealistic version of \gls{wf}. 
In \gls{wflb}, we assume that there is no signaling overhead and the perfect arrival rate is known beforehand.
Because of its unrealistic assumptions, \gls{wflb} has the best performance and it will serve as an empirical performance bound.
\footnote{The full implementation of our experiment can be found at \url{https://github.com/lethanh-96/tinyqmix-mtc}}

\subsection{System performance}
    We compare the performance of the proposed method in different simulation scenarios.
    
\subsubsection{Compare delay over time when the traffic is highly dynamic}

\begin{figure}[t]
    \centering
    \includegraphics[width=\linewidth]{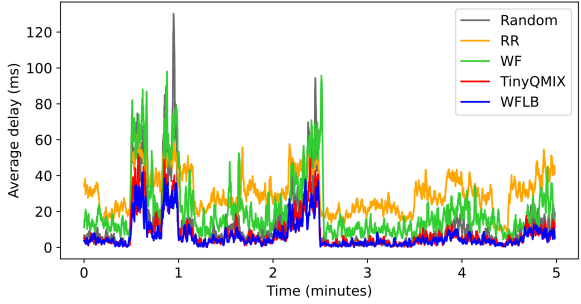}
    \caption{Moving average of delay in the first $5$ minutes of the testing trace, with $\Delta T = 10$s and $N'=12$.}
    \label{fig:evaluation:delay-time-dynamic}
\end{figure}

Figure \ref{fig:evaluation:delay-time-dynamic} compares the delay of random, \gls{rr}, \gls{wf} and TinyQMIX policy under the scenario that traffic condition changes every 10 seconds.
The delay of \gls{wf} was among the highest because $6/50$ timeslots are reserved for downlink resource assignment, while these timeslots can be used for uplink data in other distributed methods.
The performance of \gls{rr} was far from the lower-bound performance of \gls{wflb}.
Although \gls{rr} does not have any signaling overhead like \gls{wf}, its average delay still fluctuated around 40 ms, because the arrived packets need to wait for a long time until the scheduled timeslot.

TinyQMIX consistently outperformed other policies.
Its delay approached the lower bound method \gls{wflb} throughout 5 minutes testing trace, in either low traffic intensity or high traffic intensity situation.

\subsubsection{Compare delay over cluster size}

\begin{figure*}[t]
    \centering
    \begin{subfigure}[b]{0.47\linewidth}
        \centering
        \label{fig:delay:node}
        \includegraphics[width=\linewidth]{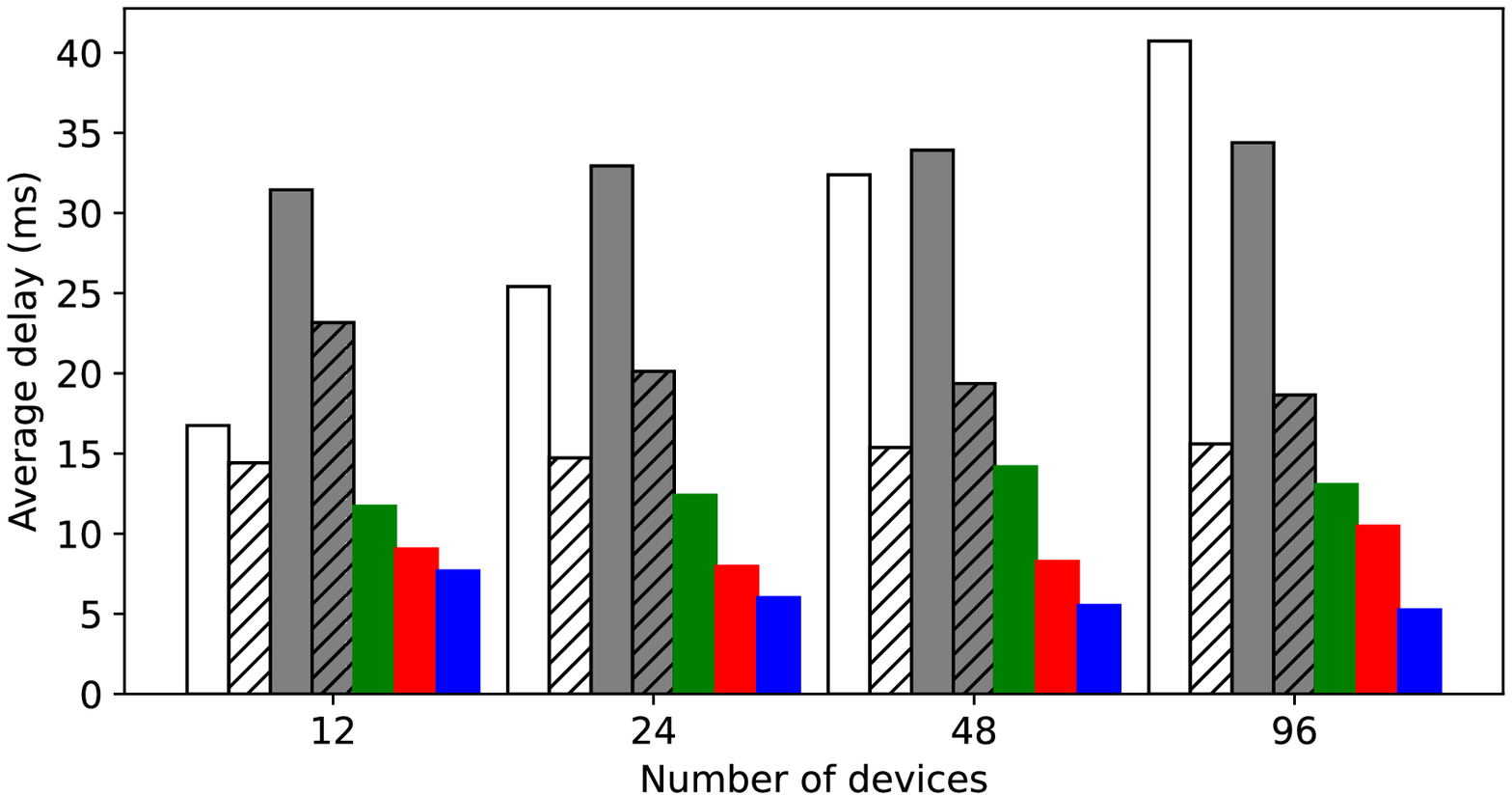}
        \caption{Average access delay with different numbers of devices per cluster, when the traffic changes every 10 seconds.}
    \end{subfigure}
    \begin{subfigure}[b]{0.47\linewidth}
        \centering
        \label{fig:delay:dynamic}
        \includegraphics[width=\linewidth]{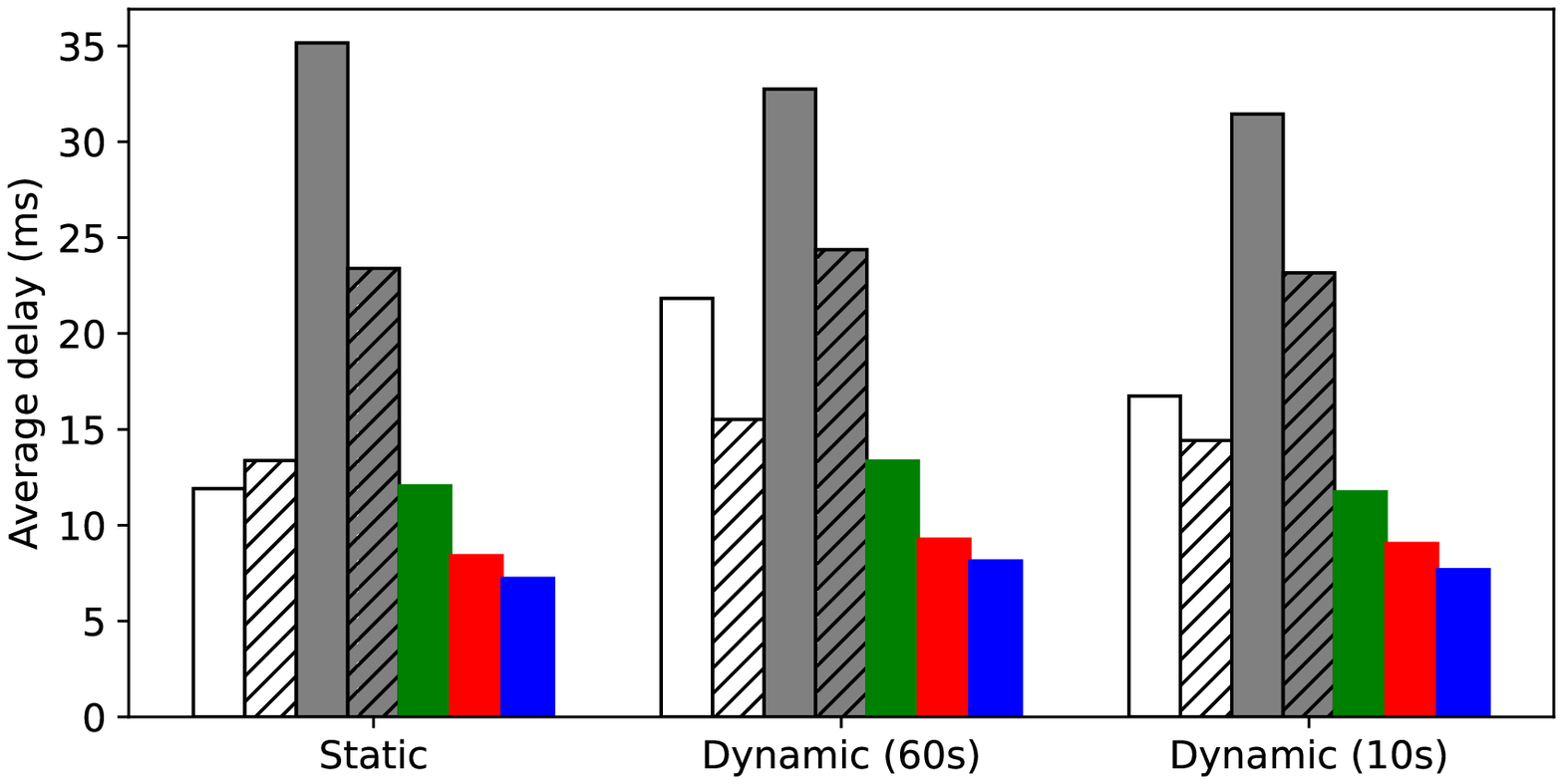}
        \caption{Average access delay with different intervals of traffic changes, when the number of nodes per cluster is 12.}
    \end{subfigure}
    \begin{subfigure}[b]{\linewidth}
        \centering
        \includegraphics[width=0.76\linewidth]{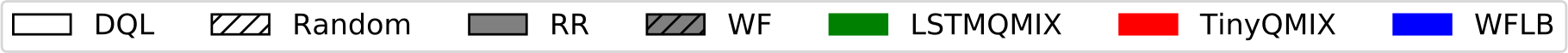}
        \label{fig:delay:legend}
    \end{subfigure}
    \caption{Average access delay under different network sizes and traffic changing rate}
\end{figure*}
Figure 3a presents the average delay of different policies when the number of devices per cluster increases.
TinyQMIX consistently outperformed the other policies.
As the cluster scaling up, \gls{ddql} and LSTMQMIX became worse, while \gls{wf} improved slightly but all of them were not outperform TinyQMIX.
It shows that TinyQMIX can consistently handle the task of distributed resource unit selection better than the baselines on all tested network scales.
When the size of the cluster increases, the delay of \gls{wflb} became smaller because there is higher flexibility for selecting resource units.
However, the joint-action space of multiple agents also became exponentially larger, which makes training the TinyQMIX harder.
The results suggest that the best cluster size is $24$, which TinyQMIX can produce the lowest delay, in comparison with other cluster sizes.

\subsubsection{Compare delay over different traffic dynamic scenarios}

Figure 3b compares different policies under three traffic dynamic scenarios.
We are able to reproduce the result from \cite{liu2021distributed}, that is \gls{ddql} is better than random and \gls{rr} policies under the static traffic trace.
However, the average delay of \gls{ddql} was higher than a random policy at a higher traffic dynamic, which indicates that \gls{ddql} cannot handle dynamic traffic as we hypothesized.
Besides, there is no significant change in the delay produced by \gls{rr}, random, or \gls{wf} policies in all three testing traces.

LSTMQMIX and TinyQMIX were trained on the scenario in which $\Delta T=10$s and tested on slower traffic changes.
LSTMQMIX only performed well on the trained scenario, while TinyQMIX performed well in all cases.
This indicates that TinyQMIX generalized the trained environment better than LSTMQMIX.
In all traces, the gap between the delay of TinyQMIX and \gls{wflb} was always the smallest.
These results suggest that the proposed TinyQMIX policy is the most suitable policy for tackling both static and highly dynamic traffic.

\subsection{Model complexity}

Here, we compare the model complexity of different fully distributed policies.
Note that \gls{ddql} or \gls{wf} requires regular information exchange between \gls{bs} and devices, thereby comparing the model complexity of fully uncoordinated policy such as TinyQMIX with \gls{ddql} or \gls{wf} is unfair.
We compare fully distributed policies such as Random, TinyQMIX, and LSTMQMIX in terms of their \gls{flops} needed to compute the local observation and perform model inference.
Random has the smallest computational requirement at only 40 \gls{flops} for selecting 40 random actions per second.
On the other hand, TinyQMIX consumes 3000 \gls{flops} when $N'=12$, 4520 \gls{flops} when $N'=24$, and just below 50k \gls{flops} when $N'=96$.
LSTMQMIX is the most demanding agent, which requires at least 52k \gls{flops} for the smallest subgroup size of 12.
Note that, common general purpose microprocessor such as ARM Cortex-M has the maximum computational capacity of 1.6 GFLOPS, thereby it can clearly support TinyQMIX with the best subgroup size $N'=24$.
In short, TinyQMIX can facilitate a smaller delay than a simple method like Random, and induces significantly less computational overhead than the recently proposed \gls{madrl} policy such as LSTMQMIX.

    \section{Discussion \& Concluding Remarks}
To sum up, our work proposed TinyQMIX, a lightweight cooperative multi-agent reinforcement learning policy that enables distributed and autonomous \gls{mmtc} network.
The findings of this study support the idea that a proper distributed resource allocation method can outperform a centralized method, due to the cost of exchanging information with a centralized controller.
Besides, our results support the hypothesis that reinforcement learning, which is learned on various patterns, can generalize and adapt to changes.

\printbibliography
\end{document}